\documentstyle[prb,preprint,eqsecnum,aps]{revtex}
\tighten
\begin{document}
\draft

\title{Stability of the hard-sphere icosahedral quasilattice}

\author{H. M. Cataldo}

\address{Departamento de F\'{\i}sica\\ Facultad de Ciencias Exactas y
Naturales\\ Universidad de Buenos Aires\\
1428 Buenos Aires, Argentina}

\author{C. F. Tejero}

\address{Facultad de Ciencias F\'{\i}sicas\\ Universidad Complutense
de Madrid\\
E-28040 Madrid, Spain}

\maketitle

\begin{abstract}
The stability of the hard-sphere icosahedral quasilattice is analyzed
using the differential formulation of the generalized effective liquid
approximation. We find that the icosahedral quasilattice is metastable
with respect to the hard-sphere crystal structures. Our results agree
with recent findings by McCarley and Ashcroft [Phys. Rev. B {\bf 49},
15600 (1994)]
carried out using the modified weighted density approximation.

\end{abstract}


\narrowtext

\section{Introduction}
\label{sec1}
The observation in 1984 by Shechtman et al\cite{Shechtman} of a sharp
diffraction pattern in a AlMn alloy with
the symmetry of the icosahedron, opened a new
field in condensed matter physics. Ever since experimental
evidence of other materials having sharp diffraction patterns with
symmetries forbidden by classical crystallography has continued to
grow.
The name quasicrystals has been coined
to represent systems with perfect order but without periodicity, i.e.
quasiperiodic systems.

Since the nonperiodic three-dimensional ($3D$) Penrose tiling
\cite{Elser}
has a diffraction pattern closely similar to that of icosahedral
alloys, it
has been extensively studied to account for icosahedral point symmetry
and also because of its relative simplicity. The $3D$
Penrose tiling is usually constructed by projection from
a $6D$ simple cubic lattice.
The projection is performed by first
defining an acceptance domain in the $3D$ complementary space in
order to select
what points of the $6D$ simple cubic lattice are effectively projected
to form the $3D$ quasilattice.
In the present investigation the  $3D$ Penrose tiling
has been generated by a special
choice of the shape and size of the acceptance domain as
described by Elser\cite{Elser}.
The last step to model a quasicrystal concerns the decoration i.e., the
location of the lattice points forming the quasicrystal and a choice
for
the pair interaction potential.

Of the many questions about quasicrystals, one concerns the stability
of
these phases.
The first theoretical approaches to such a question\cite{mermin,kalu}
were based upon the Landau theory of crystallization where the free
energy is expanded in powers of an order parameter related to density
waves with icosahedral symmetry. As a main result it was shown that
multi-component systems are required to achieve stability, a fact that
agrees with the experimental findings\cite{mermin}.

On the other hand and from a more microscopic viewpoint,
 the analysis of the stability of quasicrystals may be
rather difficult on general
grounds, but an important simplification occurs if a hard-sphere pair
potential is assumed. Indeed, a simple calculation of the maximum
packing fraction, i.e. the fraction of the total volume occupied by
the
spheres, provides an important criterion of the stability chances. For
instance, a $3D$ Penrose tiling with all vertices occupied by
identical
hard spheres leads to a fluid-like packing fraction and thus it must
certainly be
discarded as a model of a quasicrystal. Moreover, the interest of
considering hard-sphere quasicrystals goes far beyond simplicity.
Thus,
numerical studies\cite{Roth} have shown that the crucial criterion for
the quasicrystal stability with
more realistic, e.g., Lennard-Jones, interactions is the
packing fraction of the quasicrystalline hard-sphere decoration.

There are basically two ways of improving the poor packing fraction of
 the
above fully occupied Penrose tiling. These options are to change the
decoration\cite{Henley} or the acceptance domain\cite{Oguey}. Both
procedures
give approximately the same optimal packing fraction $\simeq 0.63$, a
value
that now indeed justifies a further stability study. Recently,
McCarley
and Ashcroft\cite{Ash} have studied the hard-sphere
quasicrystal using a modification of
the acceptance domain\cite{Oguey}, to obtain from the modified
weighted density
approximation a metastable quasicrystal with respect to the
crystalline and fluid phases. Their method is entirely formulated in
the $6D$ reciprocal
space which avoids direct summations over the quasilattice.
However, a drawback of this method is that a truncation of the
sum in the $6D$ reciprocal lattice is needed,
the induced estimate error in the free energy per particle of the
quasilattice being $\simeq 2\%$.

In the present paper we will consider the above optimal hard-sphere
decoration of the Penrose tiling. Our treatment is based on
the generalized effective liquid
approximation\cite{Lutsko}
which has been previously applied to perfect hard spheres and hard
disks crystals
yielding very accurate results as compared to the simulation data.
The quasilattice sums are calculated in the $3D$ real
space using a method which substantially improves convergence errors
in
comparison to previous $6D$ reciprocal lattice treatments.

The paper is organized as follows. In section \ref{sec2} we briefly
review
the generation of the $3D$ Penrose tiling which allows us to introduce
the
hard-sphere decoration. Section \ref{sec3} summarizes the generalized
effective liquid approximation for the determination
of the free energy of the quasicrystal.
Our results are presented in section \ref{sec4} together with a
discussion concerning the evaluation of quasilattice sums, while in
the
final section \ref{sec5} we gather our conclusions.

\section{The icosahedral quasilattice}
\label{sec2}

In what follows, we consider the icosahedral quasilattice obtained by
projecting a subset (to be specified below) of the
$6D$  simple cubic lattice onto
a $3D$ hyperplane\cite{Elser} $X_{{\scriptscriptstyle\parallel}}$. The
orientation of
$X_{{\scriptscriptstyle\parallel}}$
relative to the lattice is determined by requiring that the
projected vectors $\{\mbox{{\bf e}}_{{\scriptscriptstyle\parallel}}^j
\}$ of the
$6D$ basis vectors
$\{\mbox{{\bf e}}^j\}$ ($j=1,2,...6$) coincide with the
six vertex axes of the icosahedron, i.e.:

\begin{equation}
\label{21}
\mbox{{\bf e}}_{{\scriptscriptstyle\parallel}}^j=P_{{
\scriptscriptstyle
\parallel}
}^{jk}\mbox{{\bf e}}^k
\end{equation}
where a summation over repeated labels is understood. The matrix
representation of the projection operator $P_{{\scriptscriptstyle
\parallel}}$
is given by:

\begin{equation}
\label{22}
P_{{\scriptscriptstyle\parallel}} =\frac{1}{\sqrt{20}}
\left(\begin{array}{rrrrrr}
\sqrt{5} & 1 & 1 & 1 & 1 & 1 \\
1 & \sqrt{5} & 1 & -1 & -1 & 1 \\
1 & 1 & \sqrt{5} & 1 & -1 & -1 \\
1 & -1 & 1 & \sqrt{5} & 1 & -1 \\
1 & -1 & -1 & 1 & \sqrt{5} & 1 \\
1 & 1 & -1 & -1 & 1 & \sqrt{5} \\
\end{array}\right)
\end{equation}
and an elementary calculation leads to $|\mbox{{\bf e}}^j_
{\scriptscriptstyle\parallel}|^2=
1/2$ ($j=1,2,...6$) and

\[ \cos({\bf e}^1_{\scriptscriptstyle\parallel},{\bf e}^j_
{\scriptscriptstyle\parallel})=\sqrt{5}/5\;\;\;
(j=2,...,6) \]
\[ \cos({\bf e}^{2+j}_{\scriptscriptstyle\parallel},{\bf e}^{2+k}_
{\scriptscriptstyle\parallel})=\left\{ \begin{array}{rll}
  \sqrt{5}/5 & \;\;\;\;\;\;\;\; & j-k=\pm1,\, \pm4 \\
- \sqrt{5}/5 & \;\;\;\;\;\;\;\; & j-k=\pm 2,\, \pm 3
\end{array}
\right. \]
showing that the vectors $\{\mbox{{\bf e}}_{{\scriptscriptstyle
\parallel}}^j\}$
may be identified
with the six vertex directions of the icosahedron.

We also consider the $3D$ hyperplane perpendicular
to $X_{{\scriptscriptstyle\parallel}}$,
$X_{{\scriptscriptstyle\perp}}$, obtained upon projection of the
 $6D$ basis
vectors
$\{\mbox{{\bf e}}^j\}$ by the complementary projector
$P_{{\scriptscriptstyle
\perp}}$:

\begin{equation}
\label{23}
\mbox{{\bf e}}^j_{{\scriptscriptstyle\perp}}=
P_{{\scriptscriptstyle\perp}}
^{jk}\mbox{{\bf e}}^k; \,\,\,\,\,\,\,\,
P_{{\scriptscriptstyle\perp}}^{jk} = \delta^{jk}
-P_{{\scriptscriptstyle
\parallel}}^{jk}
\end{equation}
where $\delta^{jk}$ is the Kronecker delta. It can be readily
shown that
the projected vectors $\mbox{{\bf e}}_{{\scriptscriptstyle\perp}}^j$
 may also
be identified
with the six vertex directions of the icosahedron, but permuted
with respect to the projected vectors $\mbox{{\bf e}}_{
{\scriptscriptstyle\parallel}}^j$, i.e.
$\mbox{{\bf e}}_{{\scriptscriptstyle\perp}}^j\cdot\mbox{{\bf e}}_{{
\scriptscriptstyle\perp}}^k=
-\mbox{{\bf e}}_{{\scriptscriptstyle\parallel}}^j\cdot\mbox{{\bf e}}_{
{\scriptscriptstyle\parallel}}^k$ ($j\neq k$).

Both projections are dense in the $3D$ space but a
quasilattice of finite density can be constructed by projecting onto
$X_{{\scriptscriptstyle\parallel}}$ only those points whose
perpendicular space
projection lies
within a bounded region $\chi$ known as the acceptance
domain\cite{Elser}.
To construct this bounded region we
take the twenty distinct triplets
$\{\mbox{{\bf e}}_{{\scriptscriptstyle\perp}}^i,\mbox{{\bf e}}_{{
\scriptscriptstyle\perp}}^j,\mbox{{\bf
e}}_{{\scriptscriptstyle\perp}}^k\}$
of the projected vectors in $X_{{\scriptscriptstyle\perp}}$. Each
triplet
defines a
rhombohedron of volume
$v_{ijk}=|\mbox{{\bf
e}}_{{\scriptscriptstyle\perp}}^i\times\mbox{{\bf e}}_{{
\scriptscriptstyle
\perp}}^j\cdot\mbox{{\bf e}}_{{
\scriptscriptstyle\perp}}^k|$.
It is easily found that half of these rhombohedra are ``large'', i.e.
$v_{ijk}=\sqrt{8}\,\mbox{sin}(2\pi/5)/10$
and half ``small'', i.e.
$v_{ijk}=\sqrt{8}\,\mbox{sin}(4\pi/5)/10$. The disjoint union of these
twenty rhombohedra defines a closed convex region $\chi$ named the
{\em triacontahedron}
of volume $v=\sqrt{8}\,[\mbox{sin}(2\pi/5)+\mbox{sin}(4\pi/5)]$. The
triacontahedron is therefore the projection onto $X_{{
\scriptscriptstyle\perp}}$ of the unit
cell of the $6D$ cubic lattice.

The selection of a subset of the $6D$ simple cubic lattice is
accomplished
by requiring that the orthogonal projections \{${\bf r}_{{
\scriptscriptstyle\perp}}^j\}$ of
the lattice
points \{${\bf r}^j\}$  of the $6D$ simple cubic lattice lie
within the triacontahedron\cite{nota}. This construction yields a
quasilattice in $X_{{\scriptscriptstyle\parallel}}$ which may also
be regarded
as a tiling of
the $3D$
space by two kinds of rombohedra\cite{Henley}.
Using a one-parameter Gaussian approximation for the density peaks of
the quasilattice, the one-particle density can be written as:

\begin{equation}
\label{24}
\rho({\bf r}) = \left(\frac{\alpha}{\pi}\right)^{3/2}\sum_j
W({\bf r}_{{\scriptscriptstyle\perp}}^j) \mbox{e}^{-\alpha({\bf r}
-{\bf r}_{
{\scriptscriptstyle\parallel}}^j)^2}
\end{equation}
where the sum runs over the Bravais lattice vectors of the $6D$
simple cubic
crystal, $\alpha$ is
the inverse width of the
Gaussians and the weight function $W({\bf r}_{{\scriptscriptstyle
\perp}}^j)$
is defined by:

\begin{equation}
\label{25}
W({\bf r}_{{\scriptscriptstyle\perp}}^j)=\left\{\begin{array}{cl}
1\;\;\;\;\; &
{\bf r}_{{\scriptscriptstyle\perp}}^j\in\chi\\
0\;\;\;\;\; & \mbox{otherwise}\end{array}\right.
\end{equation}

It can be shown that the first three neighbour separations
of the icosahedral quasilattice
are\cite{Henley}:  $r_1^2=(3-6\sqrt{5}/5)a^2$, $r_2=a$ and $r_3^2=
(2-2\sqrt{5}/5)a^2$ where $a=\sqrt{2}/2$. Their average coordination
numbers
are $2/\tau^2$, 6 and 6, respectively, with
$\tau$ denoting the golden ratio $\tau=(1+\sqrt{5})/2$.
By locating a hard-sphere of diameter $\sigma=r_1$ at every vertex of
the quasilattice,  the packing fraction (the fraction of the total
volume occupied by the spheres) is $\simeq 0.14$, i.e. a packing
fraction characteristic of a fluid phase. On the other hand, if we
look
for accomodating hard spheres of diameter $\sigma=a$ at every vertex,
the short distance $r_1$ does not allow this. But since the frequency
 of
the
nearest-neighbor separation is small, a better packing of hard spheres
in the quasilattice can be obtained if one of the two vertex of each
$r_1$-bond is left vacant.
These short bonds form closed rings of 10 links and chains of even or
odd links. If we also require that two adjacent sites cannot be empty
there are only two ways of placing the hard
spheres on rings and chains. In the present investigation we have
randomly located a hard sphere on a vertex of every ring and in an
endpoint of each chain. This determines
the accomodation of the remaining hard spheres, the effect of taking
different  initial localizations having a negligible effect on our
results as the number of vertex of the quasilattice increases. It can
be shown\cite{Henley} that with this procedure,
the volume occupied by the hard spheres increases
leading to a hard-sphere packing
fraction
$\eta\simeq 0.629$, which is close to the random packing fraction
($\simeq 0.64$) and below the packing fraction of the crystal
structures
($\simeq 0.74$ and $\simeq 0.68$ for the fcc and bcc crystals,
respectively). This is, to our knowledge, the best icosahedral
 packing
fraction of identical hard spheres reported so far.

\section{Free energy of the hard-sphere icosahedral quasilattice}
\label{sec3}

In recent years, the freezing of hard spheres into perfect crystals
has been successfully described
by several nonperturbative density functional theories.
We here
consider the stability of the hard-sphere icosahedral quasilattice
described in \ref{sec2} within one of such approaches, the
generalized
effective liquid approximation, which is now briefly summarized.

The Helmholtz free energy  $F$ of a solid
characterized by a one-particle density $\rho({\bf r})$ is a
functional of $\rho({\bf r})$, denoted by $F=F[\rho]$, which can be
split as $F[\rho] =F_{\mbox{id}}[\rho]+F_{\mbox{ex}}[\rho]$ where

\begin{equation}
\label{31}
\beta F_{\mbox{id}}[\rho]  =  \int\,d{\bf r}\rho({\bf r})
[\ln\{\Lambda^3\rho({\bf r})\}-1]
\end{equation}
is the ideal contribution with $\beta=1/k_BT$  the inverse temperature
and $\Lambda$ the thermal de Broglie wavelength and
\begin{equation}
\label{32}
\beta F_{\mbox{ex}}[\rho] = -  \int\,d{\bf r}\rho({\bf r})\,\int\,
d{\bf
r}^{\prime} \rho({\bf
r}^{\prime})\,\int_0^1\,d\lambda\,(1-\lambda)\,
c({\bf r},{\bf r}^{\prime};[\lambda\rho])
\end{equation}
is the excess term.
In (\ref{32}) $c({\bf r},{\bf r}^{\prime};[\lambda\rho])$
is the direct  correlation function of the solid and
$\lambda$ ($0\leq\lambda\leq 1$)
is a parameter defining a linear path of integration in the space of
density functions $\rho_{\lambda}({\bf r})=\lambda\rho({\bf r})$
connecting a zero reference density to the one-particle
density $\rho({\bf r})$ of the solid.
The equilibrium solid density $\rho({\bf r})$,
determined by functional differentiation, is the
minimimum value of $F[\rho]$ at constant average density.
This variational calculation
implies the direct correlation function of the solid which is the only
unknown in (\ref{31}-\ref{32}) and hence some explicit approximations
for $F_{\mbox{ex}}[\rho]$ are required.

Based on the similarity of the thermodynamic properties of the
solid and fluid phases, the generalized effective liquid
approximation first maps the excess free energy per particle of the
solid onto that of some effective liquid, i.e.:

\begin{equation}
\label{33}
\frac{1}{N} \int\,d{\bf r}\rho({\bf r})\,\int\,d{\bf
r}^{\prime} \rho({\bf
r}^{\prime})\,\int_0^1\,d\lambda\,(1-\lambda)\,
c({\bf r},{\bf r}^{\prime};[\lambda\rho]) =
\hat{\rho}\,\int\,d{\bf r}\,\int_0^1\,d\lambda\,(1-\lambda)\,
c(|{\bf r}|;\lambda\hat{\rho})
\end{equation}
where $\hat{\rho}$ is the density of the effective liquid,
$N=\int\,d{\bf r}\rho({\bf r})$ is the number of particles, and
$c(|{\bf r}|;\lambda\hat{\rho})$
is the direct correlation function of the
liquid.
Equation (\ref{33}) is referred to as the thermodynamic mapping.

In a second step, the generalized effective liquid approximation
defines
an structural mapping in which the direct correlation function of the
solid is mapped onto that of a liquid. However, this mapping
cannot be done directly because the direct correlation function of the
liquid is translationally invariant while that of the solid
is not. But taking into account that in
(\ref{32}) the direct correlation function of the solid appears
doubly weighted by the solid density , the difficulty is
overcome by defining the structural mapping as:

\begin{equation}
\label{34}
\int\,d{\bf r}\rho({\bf r})\,\int\,d{\bf
r}^{\prime} \rho({\bf
r}^{\prime})\,
c({\bf r},{\bf r}^{\prime};[\rho]) =
\int\,d{\bf r}\rho({\bf r})\,\int\,d{\bf
r}^{\prime} \rho({\bf
r}^{\prime})\,
c(|{\bf r}-{\bf r}^{\prime}|;\hat{\rho}[\rho])
\end{equation}

With (\ref{32}-\ref{34}) a self-consistent non-linear integral
equation
is obtained for the determination of $\hat{\rho}[\rho]$ in terms of
$\rho({\bf
r})$ and the direct correlation function of the liquid. The
complicated functional
dependence $\hat{\rho}[\rho]$ can be simplified if $\rho({\bf r})$
is described in terms of a single order parameter $\alpha$ as in
(\ref{24})
in which case $\hat{\rho}$ becomes an ordinary function of $\alpha$.
The equilibrium solid density (i.e. $\alpha$) is
then determined by minimizing at constant average density the solid
free
energy with respect to the Gaussian width parameter $\alpha$ for a
given
crystal structure.

As explained elsewhere\cite{Tejero},
the non-linear integral equation for the
determination of $\hat{\rho}[\rho]$ can be further transformed
into a system of two coupled nonlinear differential
equations in $\hat{\eta}(\lambda)$:

\begin{equation}
\label{35}
\hat{\eta}^{\prime}(\lambda)
= \frac{z(\lambda)-\psi(\hat{\eta}(\lambda))}
{\lambda\psi^{\prime}(\hat{\eta}(\lambda))}
\end{equation}
and $z(\lambda)$:
\begin{equation}
\label{36}
 z^{\prime}(\lambda)=\Phi(\hat{\eta}(\lambda))
\end{equation}
where $\hat{\eta}(\lambda) = \pi\hat{\rho}(\lambda)\sigma^3/6$ is the
effective liquid packing fraction and $\sigma$ is the hard-sphere
diameter.

Using the one-parameter approximation (\ref{24}) for the one-particle
density of the quasilattice, $\Phi(\hat{\eta}(\lambda))$ is given by:

\begin{equation}
\label{37}
\Phi(\hat{\eta}(\lambda))= -\frac{1}{N}
\sum_{i=1}^{N}\sum_{j=1}^{N}\,W({\bf r}_{{\scriptscriptstyle
\perp}}^i)\,W({\bf r}_{{\scriptscriptstyle\perp}}^j)
\int_0^{\infty}\, dR\,R\, c(R;\hat{\eta}(\lambda)) S(R;\alpha,r_{ij})
\end{equation}
where  $r_{ij}= |{\bf r}_{{\scriptscriptstyle\parallel}}^i-{\bf r}_{
{\scriptscriptstyle\parallel}}^j|$ and

\begin{equation}
\label{38}
S(R;\alpha,r_{ij}) =\left[\frac{\alpha}{2\pi r_{ij}^2}\right]^{1/2}
\left[ \exp(-\alpha(R-r_{ij})^2/2)-\exp(-\alpha(R+r_{ij})^2/2)\right]
\end{equation}

In (\ref{35}) and (\ref{36}) the
prime denotes the derivative with respect to the argument
and $\psi(\hat{\rho})/\beta$
is the excess free energy per particle of the fluid
phase. For the latter we will use the Carnahan-Starling
compressibility
factor to obtain $\psi$ by thermodynamic integration of the
equation of state while the Percus-Yevick
equation is used for the structure of the fluid phase, i.e.
 the direct
correlation function.
Eqs. (\ref{35}) and (\ref{36}) have to be integrated numerically from
$\lambda=0$ to $\lambda=1$ with
initial conditions $\hat{\eta}(0) =z(0)=0$ and the excess free energy
per
particle of the quasilattice (\ref{32}) is finally determined as
$\psi(\hat{\eta}(1))$.

\section{Results}
\label{sec4}
Before looking for a numerical solution of (\ref{35}-\ref{36}),
we deal with a delicate point concerning the convergence of
the quasilattice sums in (\ref{37}). In order to emphasize it
let us rewrite the rigth hand side of
eq. (\ref{37}) in the  form:

\begin{equation}
\label{41}
\frac{1}{N}\sum_{i=1}^{N}\sum_{j=1}^{N}A(r_{ij})
\end{equation}

We first note that
for a Bravais lattice (\ref{41}) reduces to,

\begin{equation}
\label{42}
\sum_{j}\, z_j A(r_j)
\end{equation}
where the sum runs over spherical shells of sites centered around the
site at the origin, $r_j$ is the distance of shell $j$ to the origin,
and $z_j$ is the number
of sites at the $jth$-shell. In passing from (\ref{41}) to (\ref{42})
the translational symmetry of the Bravais lattice has been used.
Moreover, since $A(r_j)$ decreases rapidly with
distance, only a relatively small number of shells around the origin
give a nonnegligible contribution to the sum
leading to a rapid convergence of (\ref{42}).

But if the lattice does not have translational symmetry,
the evaluation of (\ref{41}) becomes a delicate numerical problem if
one
looks for achieving a rapid convergence.
For instance, if we consider a quasilattice of $N$
sites the
convergence of (\ref{41}) (resulting from considering the $N(N-1)/2$
different pairs $i\neq j$) becomes so slow as $N$ increases
that it remains unreachable through usual computational efforts. This
is also the case for a Bravais lattice as it may be tested by
evaluating
the sum (\ref{41}) which, on the other hand, can be easily
determined through (\ref{42}).

As stated above, a possible way for dealing with the
quasilattice sum\cite{Ash} is
to use the $6D$ reciprocal lattice. However, this procedure
only provides a partial solution to the convergence of the sum since
it
is necessary to truncate the sum at some maximum value of the
reciprocal
lattice vector leading to an
estimate error in the free energies of about $2\%$.

We here propose an alternative solution to the convergence problem
of (\ref{41}) which provides a substantial
reduction of errors and computation time in the determination of the
quasicrystal free energy. By starting with the $6D$ simple cubic
lattice, we
construct the quasilattice using the projection formalism described
in
\ref{sec2}. Let $M>>1$ be the number of the lattice points generated
in
the quasilattice
and draw an spherical surface containing almost all the
lattice points. Inside the sphere we construct a concentric
sphere with $N>>1$ ($N<M$) lattice points.

By rewriting (\ref{41}) as:

\begin{equation}
\label{43}
A(0) + \frac{1}{N}\sum_{i=1}^{N}\left[\sum_{j\neq i}A(r_{ij})\right]
\end{equation}
where we have separated the $r_{ij}=0$ contributions, we calculate the
the $i\neq j$-terms by first choosing a lattice point $i$ ($i =
1,2,... N$) inside the sphere and
then summing consecutively over all neighbouring $j$ lattice points
until the
relative error of the sum in brackets in (\ref{43}) is less than a
prefixed value. For
lattice points $i$ well inside the sphere, our procedure takes into
account all the relevant ``interactions'' in (\ref{41}). However,
if the lattice point $i$ lies near the surface of the sphere,
our procedure overestimates the ``interactions'' in (\ref{41}) because
the sum in brackets contains points outside the sphere. We have found
that
these boundary errors can be reduced by increasing $N$ (and therefore
$M$), in such a way that for $M = 22000$ and $N = 10000$, the estimate
error of the free energy per particle of the quasilattice is about
$0.1\%$.
Under such conditions, the computation time needed for evaluating the
variational free energy (for each pair $\alpha$-$\eta$) is around $20'
$
c.p.u. in a VAX 9000.

In Fig. 1, the variational free energy per particle of the
quasicrystal
$\beta\phi=\beta F/N-3\ln(\Lambda/\sigma)+1$ is represented
versus $Y=(\alpha\sigma^2/2)^{1/2}$
for different packing fractions. We have found minima for the
quasicrystal free energy as a function of the Gaussian width parameter
$\alpha$ for $\eta\geq 0.51$, i.e. a stable or metastable
quasicrystalline phase.
Similarly to ref.\cite{Ash} we have found that the quasicrystal turns
out to be the more localized phase since the minima are always
situated at greater $Y$-values than the compact fcc crystalline
phase. Our results are gathered in Table \ref{tab1}
where the free energy and the Gaussian width parameter at the
free energy mimimum of the quasicrystal
are compared to those of the fcc crystal.

In Fig. 2 we represent the solid free energy per particle versus the
packing fraction $\eta$ for
the three solid phases (fcc, bcc and quasicrystal).
We also include in the figure the fluid free energy obtained by
thermodynamic integration of the Carnahan-Starling equation of state.
It is seen that the quasicrystal is always metastable with respect
to the remaining phases, the gaps of the free energy being somewhat
less than those reported by M$^{c}$Carley and Ashcroft\cite{Ash}.

\section{Conclusions}
\label{sec5}
We have analyzed the stability of a hard-sphere quasicrystal obtained
from a simple decoration of the $3D$ Penrose tiling which has been
designed for
optimizing the packing fraction.
The stability has been analyzed using the generalized effective liquid
approximation. A simple method for evaluating the
quasilattice sums in the $3D$ real space has been formulated. The
method
minimizes the boundary effects of finite quasilattices leading to a
substantially better convergence than previous works.

Our results show that the quasicrystal is metastable with
respect to the crystalline and fluid phases. Such results agree with
recent reported\cite{Ash} calculations for a hard-sphere quasicrystal
obtained from the modified weighted density approximation.
Therefore, within these nonperturbative density functional theories
entropy is insufficient to stabilize one-component quasicrystals.

Since all known quasicrystals have complex metallic alloy phase
structures it has been argued that for the stability of quasicrystals
 it
is neccesary to have at least two class of atoms. The generalization
of
the one-component quasicrystal structure to an ordered two-component
structure has been investigated by M$^{c}$Carley and Ashcroft
\cite{Ash}
who concluded that small changes in the diameter ratio of the
two-component hard-sphere quasicrystal are not a stabilizing factor.
It should
be expected that energetic contributions resulting from considering
more
realistic interactions would propitiate stability. This possibility
seems to be ruled out using the well-known perturbation schemes when
applied
to quasicrystals in view of the great free energy differences between
the crystalline and quasicrystalline hard-sphere phases. Thus, the
stability of quasicrystal structures within the modern
density functional approaches is at present an open question.

\section{Acknowledgments}

We thank R. Brito and J. M. R. Parrondo for useful discussions.
H. M. Cataldo has been supported by grants PID 97/93 from
Consejo Nacional de Investigaciones Cient\'{\i}ficas y T\'ecnicas
(Argentina) and EX100 from Universidad de Buenos Aires.
C. F. Tejero acknowledges the DGICYT, Spain (PB91-0378) for its
financial support.

\begin{figure}
\caption{Variational free energy per particle of the
quasilattice $\beta\phi=\beta F/N-3\ln(\Lambda/\sigma)+1$
vs the Gaussian width parameter $Y=(\alpha\sigma^2/2)^{1/2}$
for packing
fractions $\eta=0.51-0.60$, by steps of 0.01 (from bottom to top).}
\label{fig1}
\end{figure}

\begin{figure}
\caption{Free energy per particle
$\beta\phi=\beta F/N-3\ln(\Lambda/\sigma)+1$  vs the packing fraction
$\eta$ for the solid phases: fcc (dotted line), bcc (medium dashed
line)
and
quasicrystal (dashed line), and for the fluid phase (continuous
line).}
\label{fig2} \end{figure}

\begin{table}
\caption{Free energy $\beta\phi=\beta F/N-3\ln(\Lambda/\sigma)+1$
and Gaussian width parameter
$Y= (\alpha\sigma^2/2)^{1/2}$ at the free energy minimum for
the quasicrystal $(q)$
and the fcc crystal $(c)$ phases at different packing fractions
$\eta$.
\label{tab1}}
\begin{tabular}{ccrcr}
$\eta$ & $\beta\phi_q$ & $Y_q$ & $\beta\phi_c$ & $Y_c$\\
\tableline
0.51 & 6.15 & 6.5 & 5.28 & 5.9\\
0.52 & 6.43 & 7.0 & 5.48 & 6.5\\
0.53 & 6.73 & 8.0 & 5.68 & 7.0\\
0.54 & 7.04 & 9.0 & 5.88 & 7.6\\
0.55 & 7.37 & 10.0 & 6.09 & 8.2\\
0.56 & 7.74 & 11.0 & 6.31 & 8.9\\
0.57 & 8.14 & 12.0 & 6.53 & 9.7\\
0.58 & 8.59 & 13.5 & 6.76 & 10.6\\
0.59 & 9.12 & 15.0 & 7.01 & 11.6\\
0.60 & 9.74 & 16.5 & 7.26 & 12.6\\
\end{tabular}
\end{table}

\end{document}